\documentstyle[aps,axodraw]{revtex}

\begin{document}

\draft
\title{
%
%
\noindent\hfill\hbox{\rm UM-TH-98-15}\vskip 10pt
%
%
Complete 2-loop Quantum Electrodynamic Contributions\\
       to the Muon Lifetime in the Fermi Model}
\author{Timo van Ritbergen\footnote{Present address:
 Institut f\"ur Theoretische Teilchenphysik, Universit\"at Karlsruhe,
            D-76128 Karlsruhe, Germany.} \ \ and\ \ Robin G. Stuart}
\address{Randall Laboratory of Physics, University of Michigan,
                  Ann Arbor, MI 48109-1120, USA}
\maketitle
\begin{abstract}
The complete 2-loop quantum electrodynamic corrections to the muon
lifetime are calculated in the Fermi theory.
The exact result for the effects of virtual and real photons,
virtual electrons, muons as well as $e^+e^-$ pair creation is
\[
\Delta\Gamma_{{\rm QED}}^{(2)}=
      \Gamma_0\left(\frac{\alpha}{\pi}\right)^2
        \left(\frac{156815}{5184}
             -\frac{1036}{27}\zeta(2)
                         -\frac{895}{36}\zeta(3)
                         +\frac{67}{8}\zeta(4)
                         +53\zeta(2)\ln(2)  \right)
      =\Gamma_0\left(\frac{\alpha}{\pi}\right)^2
      6.743
 \]
where $\Gamma_0$ is the tree-level width.
The theoretical error in the value of the Fermi coupling constant, $G_F$,
is now rendered negligible compared to the experimental uncertainty
coming from the measurement of the muon lifetime. The overall error
in $G_F$ is then roughly halved giving
\[
G_F=(1.16637\pm0.00001)\times 10^{-5}\,{\rm GeV}^{-2}.
\]
\end{abstract}


\section{Introduction}

The Fermi coupling constant, $G_F$, plays a key r\^ole in precision tests
of the Standard Model of electroweak interactions. Along with the
electromagnetic coupling constant, $\alpha$, and the $Z$ boson mass,
$M_Z$, it is one of the best measured quantities of electroweak physics and
as such is used as input in all higher-order calculations.  $G_F$ is one
of the few quantities that is sensitive to physics at very high
energy scales and is intimately related to the
$\rho$ parameter \cite{Veltman}. It was the value of $G_F$ that
provided some of the strongest constraints on the mass of the
top quark before it was directly observed.

$G_F$ is extracted from measurements of the muon lifetime,
$\tau_\mu\equiv\Gamma_\mu^{-1}$,
which is a purely leptonic process and therefore very clean both
experimentally and theoretically. Its quoted error is
$\delta G_F/G_F = 1.7\times10^{-5}$ of which
$0.9\times10^{-5}$ is experimental and $1.5\times10^{-5}$ is theoretical;
the latter being an estimate of the size of the 2-loop corrections.
Experiments are under consideration at Brookhaven National Laboratory,
the Paul Scherrer Institute,
and the Rutherford-Appleton Laboratory that could lead to a reduction
in the experimental error on the $\tau_\mu$ of a factor of 10 or more.

The radiative corrections to muon decay in the full Standard Model
naturally factorize into two pieces \cite{Sirlinsplit}, one of which,
to a very high degree accuracy, is just the quantum electrodynamic
(QED) radiative corrections in the Fermi theory. The other piece
is left free of infrared singular contributions.
It contains purely weak corrections that can be absorbed into $G_F$ which
then possesses an enriched sample of weak sector physics.
Such a separation between QED and weak corrections is not generally
possible for charged current processes.

The 1-loop QED contributions to the muon lifetime were first calculated
over 40 years ago by Kinoshita and Sirlin \cite{KinoSirl}
and by Berman \cite{Berman}.
It is known \cite{BermSirl} that the Fermi theory in the presence
of QED is finite to leading order in
$G_F$ and to all orders in the electromagnetic coupling
constant, $\alpha$. This remarkable fact means that $G_F$ can be
defined in a physically unambiguous manner at least up to the point
where finite $W$ propagator effects begin to appear.

In this article the 2-loop QED radiative corrections to muon lifetime are
calculated in the Fermi theory. The result is used to extract an improved
value for $G_F$ in which the error is entirely due to the experimental
uncertainty.

\section{The Fermi Coupling Constant}

The Fermi theory Lagrangian, relevant for the calculation
of the muon lifetime is
\begin{equation}
{\cal L}_F={\cal L}_{{\rm QED}}^0+{\cal L}_{{\rm QCD}}^0
                  +{\cal L}_W
\label{eq:FullLagrange}
\end{equation}
Here ${\cal L}_W$ is the Fermi contact interaction
\begin{equation}
{\cal L}_W=-2\sqrt{2} G_F
    \big[\bar\psi_{\nu_\mu}^0\gamma_\lambda\gamma_L\psi_\mu^0\big].
    \big[\bar\psi_e^0\gamma_\lambda\gamma_L\psi_{\nu_e}^0\big]
\label{eq:FermiLagrange}
\end{equation}
in which $\psi_\mu$, $\psi_e$, $\psi_{\nu_\mu}$ and $\psi_{\nu_e}$
are the wave functions for the muon, the electron
and their associated neutrinos respectively. The Euclidean metric in
which time-like momenta squared are negative is used.
${\cal L}_{{\rm QCD}}^0$ is the bare Quantum Chromodynamic (QCD)
Lagrangian responsible for the strong interactions and
${\cal L}_{{\rm QED}}^0$ is the usual bare Lagrangian of QED,
\begin{equation}
{\cal L}_{{\rm QED}}^0= -\sum_f\bar\psi_f^0(ip\hskip -4pt/+m_f)\psi_f^0
-\frac{1}{4}\left(\partial_\rho A_\sigma^0
                 -\partial_\sigma A_\rho^0\right)^2
+ie^0\sum_f Q_f\bar\psi_f^0\gamma_\rho\psi_f^0 A_\rho^0.
\end{equation}
The sum is over all fermion species, $f$, with mass,
$m_f$, and electric charge, $Q_f$.
$A_\rho$ is the photon field and $\gamma_L=\frac{1}{2}(1+\gamma_5)$
denotes the usual Dirac left-hand projection operator.
The superscript ${}^0$ indicates
bare, as opposed to renormalized, quantities. For the present purposes
$G_F$ goes unrenormalized. Throughout this article
dimensional regularization \cite{dimreg} is used
for the ultraviolet (UV) divergences. The appearance of infrared
(IR) divergences is largely avoided by the methods employed here.

The formula obtained for $\tau_\mu$ by means of the
${\cal L}_{{\rm F}}$ is finite to leading order in
$G_F$ and all orders in the renormalized electromagnetic
coupling constant, $\alpha_r=e_r^2/(4\pi)$ \cite{BermSirl}.
This follows from the fact that under a Fierz rearrangement that
interchanges the wavefunctions $\bar\psi_e$ and $\bar\psi_{\nu_\mu}$
in ${\cal L}_W$ the currents remain purely left-handed vector currents.
This is in sharp contrast to the case of neutron decay in which
scalar and pseudoscalar terms are generated and for which the following
arguments break down. The radiative corrections in that case are not
finite. Considering the vector part,
$\bar\psi_e\gamma_\mu\psi_\mu$, of this effective $\mu$-$e$ current,
one sees that after fermion mass renormalization is performed
the remaining divergences are independent of the masses and
thus cancel, as for the case of pure QED.
The QED corrections to the axial vector part may be shown to be
finite by noting that the transformations
$\psi_e\rightarrow\gamma_5\psi_e$ and
$m_e\rightarrow -m_e$ leave ${\cal L}_{{\rm QED}}$ and
${\cal L}_{{\rm QCD}}$ invariant but exchanges
$\bar\psi_e\gamma_\lambda\psi_\mu\leftrightarrow
\bar\psi_e\gamma_\lambda\gamma_5\psi_\mu$.
Thus the radiative corrections to the axial-vector part of the current
are equal to those of the vector part in the limit of $m_e=0$.
In practice then the only radiative corrections to the vector
pieces in ${\cal L}_W$ need to be calculated which avoids entirely the
problems associated with $\gamma_5$ in dimensional regularization.

To lowest order in $G_F$ the expression for the muon lifetime
calculated from ${\cal L}_F$ takes the general form
\begin{equation}
\frac{1}{\tau_\mu}\equiv\Gamma_\mu=\Gamma_0(1+\Delta q).
\label{eq:QEDcorr}
\end{equation}
where
\[
\Gamma_0=\frac{G_F^2 m_\mu^5}{192\pi^3}
\]
and $\Delta q$ encapsulates the higher order QED and QCD corrections
generated by
${\cal L}_F$ and can be expressed as a power series expansion in
the renormalized electromagnetic coupling constant, $\alpha_r$,
\begin{equation}
\Delta q=\sum_{i=0}^\infty\Delta q^{(i)}
\label{eq:DeltaqSeries}
\end{equation}
in which the index $i$ gives the power of $\alpha_r$ that appears in
$\Delta q^{(i)}$.

Assuming that the electron neutrino and muon neutrino are massless
it can be shown that
\begin{equation}
\Delta q^{(0)}=-8x-12x^2\ln x+8x^3-x^4,
        \ \ \ x=\frac{m_e^2}{m_\mu^2},
\end{equation}
that comes from phase space integrations.

The ${\cal O}(\alpha)$ corrections in $\Delta q$, first
obtained by Kinoshita and Sirlin \cite{KinoSirl} and by
Berman \cite{Berman}, are
\begin{equation}
\Delta q^{(1)}=
\left(\frac{\alpha_r}{\pi}\right)\left(\frac{25}{8}-3\zeta(2)\right)
+{\cal O}\left(\alpha_r\frac{m_e^2}{m_\mu^2}
               \ln\frac{m_e^2}{m_\mu^2}\right).
\label{eq:q1}
\end{equation}
where $\zeta$ is the Riemann zeta function and $\zeta(2)=\pi^2/6$.
An exact expression for the full electron mass dependence in
$\Delta q^{(1)}$ has been given by Nir \cite{Nir}.

Recently the hadronic contributions to $\Delta q^{(2)}$ were computed
using dispersion relations along with contributions
from muon and tau loops \cite{muonhad}. Their effect was shown
to be small relative to the present experimental error.
They become relevant for the next generation of muon
lifetime experiments but the hadronic uncertainty is still
well under control.

The Kinoshita-Lee-Nauenberg \cite{KinoshitaLeeNaue}
theorem guarantees that $\Delta q$ is free from singularities
as $m_e\rightarrow0$, other than those that can are absorbed into
$\alpha_r$. It may be shown \cite{howto} that all large logarithms of
the form
$\alpha^i\ln^{i-1}(m_\mu^2/m_e^2)$ for all $i>0$ and those of
$\alpha^3\ln(m_\mu^2/m_e^2)$ can be accounted for, in a manner
consistent both with the calculation of Ref.\cite{muonhad} and
the perturbative results presented here, by setting
\begin{equation}
\alpha_r\longrightarrow
\alpha_e(m_\mu)=\frac{\alpha}
    {1-\frac{\displaystyle \alpha}{\displaystyle 3\pi}
       \ln\frac{\displaystyle m_\mu^2}{\displaystyle m_e^2}}
                 +\frac{\alpha^3}{4\pi^2}\ln\frac{m_\mu^2}{m_e^2}
\label{eq:alphaefinal}
\end{equation}
where $\alpha$ is the experimentally-measured quantity,
$\alpha=1/137.0359895(61)$ \cite{PDG}.
The contribution to the muon lifetime
from the ${\cal O}(\alpha^2)$ logarithmic term coincides with the
result obtained in Ref.\cite{RoosSirlin}. The logarithms of
${\cal O}(\alpha^3)$ were first obtained by Jost and Luttinger
\cite{JostLuttinger}. When evaluated Eq.(\ref{eq:alphaefinal}) yields
$\alpha_e(m_\mu)=1/135.90=0.0073582$.
In the $\overline{{\rm MS}}$ renormalization scheme with 't~Hooft mass,
$\mu=m_\mu$, Eq.(\ref{eq:alphaefinal}) correctly includes non-logarithmic
terms up to ${\cal O}(\alpha^2)$ but those of ${\cal O}(\alpha^3)$
have been dropped.

\section{2-loop Corrections}
\subsection{Photonic Corrections}
\label{sec:photonloops}

The calculation of the 2-loop QED corrections to the muon lifetime
involves the sum of the cross-sections
$\mu^-\rightarrow\nu_\mu e^-\bar\nu_e$,
$\mu^-\rightarrow\nu_\mu e^-\bar\nu_e\gamma$,
$\mu^-\rightarrow\nu_\mu e^-\bar\nu_e\gamma\gamma$ and
$\mu^-\rightarrow\nu_\mu e^-\bar\nu_e e^+e^-$ with up to two
virtual photons. Individual diagrams are IR divergent and in some
cases require integration over a 5-body phase space. The problem of
cancelling these IR singularities can be avoided entirely if the QED
corrections are obtained as the imaginary part of 4-loop
propagator type Feynman diagrams by means of the optical theorem.
Some of these diagrams are shown in Fig.1.
The heavy lines represent muons which are the only particles
taken to have non-zero mass. The 4-fermion vertex used is the vector part
of the usual one from the Fermi theory. Inspection of the diagrams shows
that the cuts generating imaginary parts produce all of the Feynman
diagrams contributing to muon decay. Extra diagrams do appear in which
the cut goes through a muon line but such diagrams vanish
kinematically because the external muon is on its mass shell.

\begin{figure}
\begin{center}
\hfill
\begin{picture}(120,70)(0,0)
 \SetWidth{1.5}
 \Line(20,20)(40,20)
 \Text(20,23)[bl]{$\mu^-$}
 \Line(80,20)(120,20)
 \SetWidth{0.8}
 \ArrowLine(20,20)(40,20)
 \ArrowLine(80,20)(96,20)
 \ArrowLine(96,20)(108,20)
 \ArrowLine(108,20)(120,20)
 \SetWidth{0.5}
 \ArrowLine(40,20)(80,20)
 \Text(60.8,21)[bl]{$\nu_\mu$}
 \Oval(60,20)(16,20)(0)
 \ArrowLine(60.01,36)(59.01,36)
 \Text(63,36)[bl]{$\bar\nu_e$}
 \ArrowLine(59.99,4)(60.01,4)
 \Text(60.8,7)[bl]{$e^-$}
 \PhotonArc(80,29.2)(29.2,238,341){1.6}{11.5}
 \PhotonArc(81.6,23.2)(14,246,343){1.6}{5.5}
 \Text(70,-12)[t]{(a)}
\end{picture}
\hfill
\begin{picture}(120,70)(0,0)
 \SetWidth{1.5}
 \Line(15,20)(30,20)
 \Line(70,20)(120,20)
 \SetWidth{0.8}
 \ArrowLine(15,20)(30,20)
 \ArrowLine(70,20)(86,20)
 \ArrowLine(86,20)(104,20)
 \ArrowLine(104,20)(120,20)
 \SetWidth{0.5}
 \ArrowLine(30,20)(70,20)
 \Oval(50,20)(16,20)(0)
 \ArrowLine(50.01,36)(49.01,36)
 \ArrowLine(49.99,4)(50.01,4)
 \PhotonArc(61.1,24)(24,238,350){1.6}{10.5}
 \PhotonArc(78,31.2)(26.8,238,335){1.6}{10.5}
 \Text(70,-12)[t]{(b)}
\end{picture}
\hfill\null\\
\hfill
\begin{picture}(120,70)(0,0)
 \SetWidth{1.5}
 \Line(15,20)(30,20)
 \Line(70,20)(120,20)
 \SetWidth{0.8}
 \ArrowLine(15,20)(30,20)
 \ArrowLine(70,20)(84,20)
 \ArrowLine(84,20)(97,20)
 \ArrowLine(97,20)(107,20)
 \ArrowLine(107,20)(120,20)
 \SetWidth{0.5}
 \ArrowLine(30,20)(70,20)
 \Oval(50,20)(16,20)(0)
 \ArrowLine(50.01,36)(49.01,36)
 \ArrowLine(49.99,4)(50.01,4)
 \PhotonArc(94.8,18)(12,15,166){1.6}{6.5}
 \PhotonArc(66,36)(34,250,331){1.6}{10.5}
 \Text(70,-12)[t]{(c)}
\end{picture}
\hfill
\begin{picture}(120,70)(0,0)
 \SetWidth{1.5}
 \Line(20,20)(52.8,20)
 \Line(86.8,20)(122,20)
 \SetWidth{0.8}
 \ArrowLine(20,20)(35,20)
 \ArrowLine(35,20)(52.8,20)
 \ArrowLine(86.8,20)(106,20)
 \ArrowLine(106,20)(122,20)
 \SetWidth{0.5}
 \ArrowLine(53.2,20)(86.8,20)
 \Oval(70,20)(12,16.8)(0)
 \ArrowLine(70.01,32)(69.01,32)
 \ArrowLine(69.99,8)(70.01,8)
 \PhotonArc(60.8,24)(27.2,190,270){1.6}{7.0}
 \PhotonArc(60.8,22)(25.2,270,340){1.6}{5.5}
 \PhotonArc(79.2,22)(25.2,200,270){1.6}{6.0}
 \PhotonArc(79.2,24)(27.2,270,351){1.6}{7.5}
 \Text(70,-12)[t]{(d)}
\end{picture}
\hfill\null\\
\vglue 18pt
\end{center}
\caption{Examples of diagrams whose cuts give contributions to
        $\mu^-\rightarrow\nu_\mu e^-\bar\nu_e$,
        $\mu^-\rightarrow\nu_\mu e^-\bar\nu_e\gamma$ or
        $\mu^-\rightarrow\nu_\mu e^-\bar\nu_e\gamma\gamma$.}

\label{fig:PhotonLoopDiags}
\end{figure}
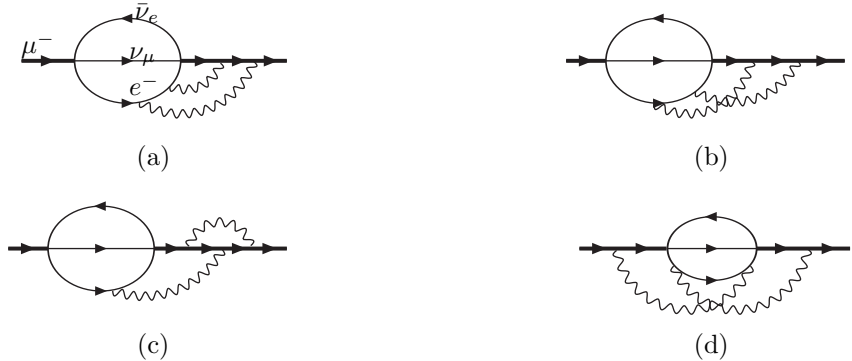

The imaginary parts of the necessary 4-loop propagator type diagrams
were calculated as follows.
Recursion relations \cite{parialint} obtained by integration-by-parts
were first applied to reduce all dimensionally regularized integrals
to a small set of relatively simple integrals.
These primitive integrals were chosen
to be free from specific IR divergences that occur on-shell.
The well-behaved primitive integrals were then calculated by
taking the external muon momentum, $q$, off mass shell
to obtain expressions as power series in $x=-q^2/m_\mu^2$ and
logarithms of $x$
using well-established large mass expansion techniques
along the lines of Ref.\cite{largemass}.
This series serves as a convenient representation
as its coefficients involve simpler integrals.
Many terms in the large mass expansion can be discarded
since they do not contribute to the imaginary part.
What remains of the coefficients in the expansion can be evaluated
in closed form in terms of polygamma functions and certain classes
of multiple nested sums \cite{sfunctions}.
Then, the on-shell limit, $x=1$, is taken and the infinite
sum over the coefficients of $x^k$ is performed.
In this process the exact expressions collapse into
known constants such as the Riemann zeta function of integer
arguments, $\zeta(k)$, and polylogarithms, ${\rm Li}_k(1/2)$.
Details of the procedures followed will be described elsewhere \cite{howto}.

Fermion mass renormalization is performed in the on-shell
scheme\footnote{That is to say that the renormalized mass of a stable
fermion is set equal to its physical or pole mass} that generates
derivatives of fermion self-energies for the external leg corrections.
All diagrams were calculated in a general covariant gauge
for the photon field and exact cancellation in the final result of the
dependence on the gauge parameter was demonstrated.

The result for just the photonic diagrams is
\begin{mathletters}
\begin{eqnarray}
\Delta\Gamma_{\gamma\gamma}^{(2)}&=&
\Gamma_0\left(\frac{\alpha_e(m_\mu)}{\pi}\right)^2
\bigg(\frac{11047}{2592}-\frac{1030}{27}\zeta(2)
                         -\frac{223}{36}\zeta(3)
                         +\frac{67}{8}\zeta(4)
                         +53\zeta(2)\ln (2)\bigg)\\
                         &=&\Gamma_0\left(\frac{\alpha_e(m_\mu)}{\pi}\right)^2
                            3.55877.
\end{eqnarray}
\end{mathletters}

\subsection{Electron Loop Corrections}
\label{sec:electronloops}

\begin{figure}
\begin{center}
\hfill
\begin{picture}(120,70)(0,0)
 \SetWidth{1.5}
 \Line(20,20)(36,20)
 \Text(20,23)[bl]{$\mu^-$}
 \Line(36,20)(52.8,20)
 \Line(87.2,20)(120,20)
 \SetWidth{0.8}
 \ArrowLine(20,20)(36,20)
 \ArrowLine(36,20)(52.8,20)
 \ArrowLine(87.2,20)(104,20)
 \ArrowLine(104,20)(120,20)
 \SetWidth{0.5}
 \ArrowArc(70,-3.2)(6.4,0,180)
 \Text(68,0)[br]{$e^+$}
 \ArrowArc(70,-3.2)(6.4,180,0)
 \Text(75,-6)[tl]{$e^-$}
 \ArrowLine(50,20)(90,20)
 \Text(70.8,20.2)[bl]{$\nu_\mu$}
 \Oval(70,20)(12,16.8)(0)
 \ArrowLine(70.01,32)(69.01,32)
 \Text(73,33)[bl]{$\bar\nu_e$}
 \ArrowLine(69.99,8)(70.01,8)
 \Text(70.8,11)[bl]{$e^-$}
 \PhotonArc(64,26)(30,191,270){1.6}{9.5}
 \PhotonArc(76,26)(30,270,349){1.6}{9.5}
 \Text(70,-17)[t]{(a)}
\end{picture}
\hfill
\hglue -0.2cm
\begin{picture}(120,70)(0,0)
 \SetWidth{1.5}
 \Line(15,20)(30,20)
 \Line(70,20)(120,20)
 \SetWidth{0.8}
 \ArrowLine(15,20)(30,20)
 \ArrowLine(70,20)(108,20)
 \ArrowLine(108,20)(120,20)
 \SetWidth{0.5}
 \ArrowArc(83.2,3.2)(6.4,0,180)
 \ArrowArc(83.2,3.2)(6.4,180,0)
 \ArrowLine(30,20)(70,20)
 \Oval(50,20)(16,20)(0)
 \ArrowLine(49.99,4)(50.01,4)
 \ArrowLine(50.01,36)(49.99,36)
 \PhotonArc(76,36)(34,242,272){1.6}{4.0}
 \PhotonArc(76,36)(34,292,330){1.6}{4.5}
 \Text(70,-17)[t]{(b)}
\end{picture}
\hfill\null\\
\hfill
\begin{picture}(120,70)(0,0)
 \SetWidth{1.5}
 \Line(20,20)(52.8,20)
 \Line(87.2,20)(120,20)
 \SetWidth{0.8}
 \ArrowLine(20,20)(52.8,20)
 \ArrowLine(87.2,20)(120,20)
 \SetWidth{0.5}
 \ArrowArc(70,-2.8)(6.4,0,180)
 \ArrowArc(70,-2.8)(6.4,180,0)
 \ArrowLine(50,20)(90,20)
 \Oval(70,20)(12,16.8)(0)
 \ArrowLine(69.99,8)(70.01,8)
 \ArrowLine(70.01,32)(69.99,32)
 \PhotonArc(68,8)(12,160,245){1.6}{4.5}
 \PhotonArc(72,8)(12,295,20){1.6}{4.5}
 \Text(70,-17)[t]{(c)}
\end{picture}
\hfill
\begin{picture}(120,70)(0,0)
 \SetWidth{1.5}
 \Line(15,20)(30,20)
 \Line(64.4,20)(125,20)
 \SetWidth{0.8}
 \ArrowLine(15,20)(30,20)
 \ArrowLine(64.4,20)(75.1,20)
 \ArrowLine(75.1,20)(109.3,20)
 \ArrowLine(109.3,20)(125,20)
 \SetWidth{0.5}
 \ArrowArc(92.2,9.3)(6.4,0,180)
 \ArrowArc(92.2,9.3)(6.4,180,0)
 \ArrowLine(30,20)(64.4,20)
 \Oval(47.2,20)(12,16.8)(0)
 \ArrowLine(47.19,8)(47.21,8)
 \ArrowLine(47.21,32)(47.19,32)
 \PhotonArc(85.8,20)(10.7,180,270){1.6}{4.5}
 \PhotonArc(98.6,20)(10.7,270,0){1.6}{4.5}
 \Text(70,-17)[t]{(d)}
\end{picture}
\hfill\null\\
\vglue 18pt
\end{center}
\caption{Diagrams containing an electron loop whose cuts give
         contributions to muon decay,
         $\mu^-\rightarrow\nu_\mu e^-\bar\nu_e$,
         $\mu^-\rightarrow\nu_\mu e^-\bar\nu_e\gamma$ or
         $\mu^-\rightarrow\nu_\mu e^-\bar\nu_e e^+e^-$.}
\label{fig:ELoopDiags}
\end{figure}
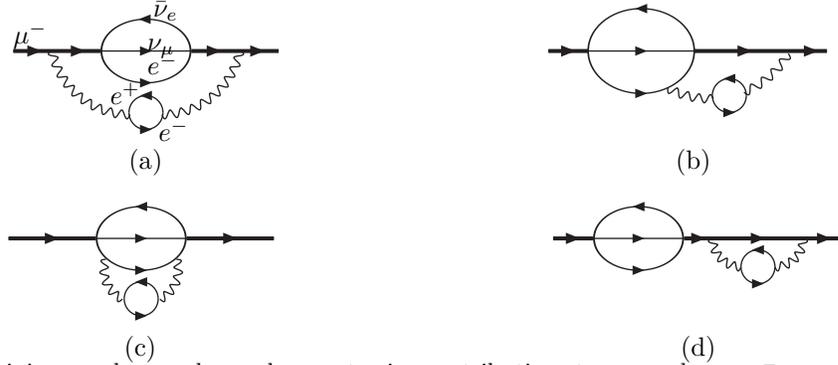

The contribution of electron loops to the muon lifetime differs from
those of other fermions in that they must be combined with diagrams
with an additional $e^+e^-$ pair in the final state in order to produce
an IR finite result, however the procedure described above
may be applied here as well.
The electron loop diagrams are shown in Fig.\ref{fig:ELoopDiags}.
To Fig.\ref{fig:ELoopDiags}d must be added a diagram
containing a muon mass counterterm, $\delta m_\mu$, on the external leg.
Furthermore, diagrams, in which the electron loop is replaced by the photon
self-energy counterterm, must be included to produce a UV finite result.
This counterterm contribution is proportional to $\Delta q^{(1)}$
and depends on the particular renormalization scheme that has
been chosen. The overall result in the
$\overline{{\rm MS}}$ renormalization scheme with 't~Hooft mass
$\mu=m_\mu$ consistent with Eq.(\ref{eq:alphaefinal}) is
\begin{mathletters}
\begin{eqnarray}
\Delta\Gamma_{{\rm elec}}^{(2)}&=&
-\Gamma_0\left(\frac{\alpha_e(m_\mu)}{\pi}\right)^2
\left(\frac{1009}{228}-\frac{77}{36}\zeta(2)
                      -\frac{8}{3}\zeta(3)\right)\\
                &=&\Gamma_0\left(\frac{\alpha_e(m_\mu)}{\pi}\right)^2
                            3.22034
\label{eq:electronnum}
\end{eqnarray}
\end{mathletters}
which is about two orders of magnitude greater than that of
either muon loops or hadrons.
The value obtained in Eq.(\ref{eq:electronnum}) is
consistent with a numerical study presented in
Ref.\cite{LukeSavaWise} in the context of semi-leptonic decays of heavy
quarks.

The same methods used to calculate the contribution from electron loops
can be applied to muon loops. Agreement was found with the result
of Ref.\cite{muonhad}.

\section{Conclusions}

The photonic corrections of section~\ref{sec:photonloops} can be
combined with those of the electron loops and $e^+e^-$ pair
production of section \ref{sec:electronloops}, and adding the
exact result for muon loops of Ref. \cite{muonhad} gives

\begin{mathletters}
\begin{eqnarray}
\Delta\Gamma_{{\rm QED}}^{(2)}&=&
      \Gamma_0\left(\frac{\alpha_r}{\pi}\right)^2
        \left(\frac{156815}{5184}
             -\frac{1036}{27}\zeta(2)
                         -\frac{895}{36}\zeta(3)
                         +\frac{67}{8}\zeta(4)
                         +53\zeta(2)\ln(2)
                          \right)\\
      &=&\Gamma_0\left(\frac{\alpha_r}{\pi}\right)^2
             6.743
\end{eqnarray}
\end{mathletters}
with $\alpha_r=\alpha_e(m_\mu)$=1/135.90.
The resulting expression contains all
corrections of ${\cal O}(\alpha^2)$,
${\cal O}\left(\alpha^3\ln(m_e^2/m_\mu^2)\right)$ and
${\cal O}\left(\alpha^i\ln^{i-1}(m_e^2/m_\mu^2)\right)$ for all $i\ge2$.
Adding the hadronic and tau loop
contributions of Ref.\ \cite{muonhad} one obtains
\begin{equation}
\Delta\Gamma^{(2)} =  \Gamma_0\left(\frac{\alpha_r}{\pi}\right)^2
                (6.700\pm 0.002).
\end{equation}
where the error is a conservative estimate of the hadronic uncertainty.
Using the current best value for
$\tau_\mu=(2.19703\pm0.00004)\,\mu$s \cite{PDG} yields
\begin{equation}
G_F=(1.16637\pm0.00001)\times 10^{-5}\,{\rm GeV}^{-2}.
\end{equation}
That represents a reduction in the overall error on $G_F$ of
about a factor of 2 and a downward shift in the central value of twice
the experimental uncertainty. $G_F$ is now known to 9\,ppm.
The next generation of measurements of the muon
lifetime are expected to reduce this by at least a further factor of 10.

\section*{Acknowledgments}

Useful discussions with and encouragement from A.~Sirlin are
gratefully acknowledged. This work was funded in part by the
US Department of Energy.


\begin{references}

\bibitem{Veltman} D. A. Ross and M. Veltman,
         Nucl.\ Phys.\ {\bf B 95} (1977) 135.

\bibitem{Sirlinsplit} A. Sirlin, Rev.\ Mod.\ Phys.\ {\bf 50} (1978) 573;
                      Phys.\ Rev.\ {\bf D 22} (1980) 971;
                      Phys.\ Rev.\ {\bf D 29} (1984) 89.

\bibitem{KinoSirl} T. Kinoshita and A. Sirlin,
        Phys.\ Rev.\ {\bf 113} (1959) 1652.

\bibitem{Berman} S. M. Berman, Phys.\ Rev.\ {\bf 112} (1958) 267.

\bibitem{BermSirl} S. M. Berman and A. Sirlin,
        Ann.\ Phys.\ {\bf 20} (1962) 20.

\bibitem{dimreg} C. G. Bollini and J. J. Giambiagi,
        Phys.\ Lett.\ {\bf 40 B} (1972) 566;\\
        G. 't Hooft and M. Veltman, Nucl. Phys. {\bf B 44} (1972) 189.

\bibitem{Nir} Y. Nir, Phys.\ Lett. {\bf B 221} (1989) 184.

\bibitem{muonhad} T. van Ritbergen and R. G. Stuart,
        Phys.\ Lett.\ {\bf B 437} (1998) 201.

\bibitem{KinoshitaLeeNaue} T. Kinoshita,
        J.\ Math.\ Phys.\ {\bf 3} (1962) 650;\\
        T. D. Lee and M. Nauenberg, Phys.\ Rev.\ {\bf 133B} (1964) 1549.

\bibitem{howto} T. van Ritbergen and R. G. Stuart, {\sl in preparation}.

\bibitem{PDG} C. Caso {\it et al.},
        Eur.\ Phys.\ J.\ {\bf C 3} (1998) 1.

\bibitem{RoosSirlin} M. Roos and A. Sirlin,
        Nucl. Phys.\ {\bf B 29} (1971) 296.


\bibitem{JostLuttinger} R. Jost and J. M. Luttinger,
        Helv.\ Phys.\ Acta {\bf 23} (1950) 201.

\bibitem{parialint}
        K. G. Chetyrkin, F. V. Tkachov, Nucl.\ Phys.\ {\bf B 192} (1981) 159;\\
        F. V. Tkachov, Phys.\ Lett.\ {\bf 100 B} (1981) 65.

\bibitem{largemass}
        S. A. Larin, T. van Ritbergen and
        J. A. M. Vermaseren, Nucl.\ Phys.\ {\bf B 438} (1995) 278.

\bibitem{sfunctions} J. A. M. Vermaseren, preprint FTUAM-98-7 (Madrid 1998),
   hep-ph/9806280

\bibitem{LukeSavaWise} M. Luke, M. J. Savage and M. B. Wise,
        Phys.\ Lett.\ {\bf B 343} (1995) 327.


\end{references}
\end{document}